 \documentclass[conference]{IEEEtran}

\IEEEoverridecommandlockouts                              

\IEEEoverridecommandlockouts

\usepackage{amsmath}
\usepackage{graphicx}
\usepackage{color}
\DeclareMathOperator{\dis}{d}
\DeclareMathOperator*{\argmax}{argmax}
\DeclareMathOperator*{\argmin}{argmin}

\begin{document}

\title{\LARGE
Classification of Traffic Using Neural Networks by Rejecting: a Novel Approach in Classifying VPN Traffic
}

\author{Ali Parchekani*, Salar Nouri*\thanks{* Both authors contributed equally to this work.}, Vahid Shah-Mansouri, Seyed Pooya Shariatpanahi
\\
School of ECE, College of Engineering,
University of Tehran,
Tehran, Iran\\
\{a.parchekani\}@mail.utoronto.ca, \{salar.nouri, vmansouri, p.shariatpanahi\}@ut.ac.ir
}

\maketitle
\thispagestyle{empty}
\pagestyle{empty}

\begin{abstract}

In this paper, we introduce a novel end-to-end traffic classification method to distinguish between traffic classes including VPN traffic in three layers of the Open Systems Interconnection (OSI) model. Classification of VPN traffic is not trivial using traditional classification approaches due to its encrypted nature. We utilize two well-known neural networks, namely multi-layer perceptron and recurrent neural network to create our cascade neural network focused on two metrics: class scores and distance from the center of the classes. Such approach combines extraction, selection, and classification functionality into a single end-to-end system to systematically learn the non-linear relationship between input and predicted performance. Therefore, we could distinguish VPN traffics from non-VPN traffics by rejecting the unrelated features of the VPN class. Moreover, we obtain the application type of non-VPN traffics at the same time. The approach is evaluated using the general traffic dataset ISCX VPN-nonVPN, and an acquired dataset. The results demonstrate the efficacy of the framework approach for encrypting traffic classification while also achieving extreme accuracy, $95$ percent, which is higher than the accuracy of the state-of-the-art models, and strong generalization capabilities. 

\end{abstract}


\section{INTRODUCTION}

Traffic classification is referred to the task of classifying traffic flows based on the class of service which indicates the application category the flow belongs to. Traffic flows are the set of packets that have the same source and destination IP and port addresses [1]. Multiple functions, including tracking, identification, control, and optimization, could then be carried out on the traffic classes [2]. To cope with the problems of increasing traffic types and transmitting speeds, researchers are pursuing lightweight algorithms with as little computing requirements as possible for classification purposes.

Virtual private networks (VPN) are employed to connect users over the internet to an enterprise network securely. VPN protects the security of information transmitted across internet using packet-level encryption. Due to the encryption of traffic, it is not very easy to carry out traffic classification for VPN connections and encrypted traffic classification which involves affiliating traffic flows towards a category of application (e.g., email, FTP) [3]. Traffic encryption methods used in VPN networks are divided into application-layer encryption, presentation-layer encryption, and network-layer encryption [4].

Different traffic classification approaches are port-based, deep packet inspection (DPI), feature-based, and host behavior-based. A port-number based approach is simple to implement and very efficient in large networks. However, some applications may not have distinct ports. DPI-based methods have several drawbacks, including significant complexity and processing load, difficulty to implement on proprietary protocols, and are not applicable to encrypted traffics. Statistical and behavioral-based techniques are essential techniques for machine learning that identify traffic by utilizing a collection of specific features of the traffic flows [5].

Wang et al. [6] proposed two machine learning based traffic classification methodologies: 1) divide and conquer model 2) end-to-end (E2E) learning model. Modifying traffic features, extracting and selecting features from raw traffic data are the first steps of traffic classification. Then, a basic machine learning classifier such as Support Vector Machine is used to classify the traffic data. These procedures are, of course, carried out independently of one another. This method is known as a divide-and-conquer approach, and it involves splitting down a major problem into multiple sub-problems [7]. Two downsides of this technique are that the optimum solution of a sub-problem does not always imply the best solution of the main problem. E2E learning, on the other hand, consists of multiple Deep Neural Network (DNN) layers and is used to overcome complicated problems. Each DNN layer can concentrate to execute intermediate processes required for such problems, similar to the human brain. Furthermore, the basic notion of E2E is that utilizing a single model that can specialize to anticipate outcome directly from inputs allows the development of otherwise highly complicated networks that may be deemed cutting-edge [8].

In this paper, we present an end-to-end approach using cascade neural network structure with rejection strategy [9] [10], for encrypted traffic classification. We employ two supervised learning-based classification algorithms, namely, multi-layer perceptron (MLP) and Long Short-Term Memory (LSTM) which is a specific recurrent neural network (RNN). Traffic classification is performed layer by layer, and elevated-level characteristics are like the activation function input. To be more precise, an encrypted E2E system of traffic classification using MLP is used to discern the most efficient form of encrypted traffic simplification. If the MLP model does not recognize the features, then the RNN approach is utilized to define the features and classify them. Eventually, we evaluate the performance of our algorithm on a public traffic dataset (i.e., ISCX VPN-nonVPN (ISCX) [3]) as well as actual data collected from a local Internet Service Provider (ISP) network. Our analysis shows substantial enhancements to the state-of-the-art approach [6].

This scenario could be employed to three different encrypted VPN in the OSI model layers: Data Link Layer, Network Layer, and Application Layer; thereby, compared to the previous state-of-the-art and machine learning methods, it is an efficient and successful method. It could also gain knowledge, especially the nonlinear relationship among the raw traffic input and the predicted performance label, rather than splitting a sophisticated issue into the meta-problems. To the best of our knowledge, this is the first paper to use a simple cascade neural network with rejection for traffic classification. The proposed model improves accuracy, precision, and recall while also reducing vulnerability to adversarial attacks [10].

The rest of the paper is organized as follows. Section II explains related work. The methodology of the suggested approach is presented in Section III. The model of the neural networks used in the paper is demonstrated in Section IV. Section V includes the results and comparison with the state-of-the-art method. Section VI involves the concluding assertions.

\section{Related Work}

To classify encrypted traffic, Bacquet et al. [11] implemented genetic programming. They used an extended multi-objective genetic algorithm feature selection and cluster count optimization for K-Means. Xie et al. [12] employed subspace clustering to instruct the current classification algorithm to classify each program independently using its related features, rather than separating one framework with the other using combined network topologies. The method demonstrated very pinpoint precision and had been versatile to adjust on five traces from various ISPs. Wright et al. [13] suggested a mechanism for morphing one traffic type to appear as something in the packet size spread, utilizing convex optimization techniques to change the packets in cleartext. Wang et al. [14] proposed a 1-dimensional convolutional neural network for E2E encrypted traffic classification. Lotfollahi et al. [15] proposed a Deep Packet framework which employed two deep neural network structures namely, stacked auto-encoder and convolutional neural network to network traffic classification. Song et. al [16] proposed a traffic classification technique based on text convolutional neural networks (T-CNN), in which traffic data is represented as vectors, and then T-CNN are used to extract necessary features for traffic classification. A simple MLP model to classify a receiving connection which uses machine learning to recognize the basic patterns of VPN and non-VPN has been proposed by Miller et al. [17]. A multimodal Deep Learning (DL) framework for mobile encrypted traffic classification is proposed by Aceto et al. [18].

Although, the mentioned articles based on machine learning approaches, have obtained good performance, they have utilized deep neural networks instead of simple neural networks to traffic classification, which raises the complexity issue.
In addition, in networks designed according to the end-to-end principle, application features reside in the network's communicating end nodes, instead of intermediate nodes, like routers, which exist to set up the network. In comparison, the classification accuracy of end-to-end approaches is somewhat weak. In some of them, features apart from actual traffic were utilized as inputs by hand which means their techniques are not fully end-to-end processes.

The implemented end-to-end encrypted traffic classification system may exclude conventional measures such as feature engineering, extraction features, and features selection that are widely employed in conventional dividing and conquering methods.

\section{Methodology}

\subsection{Data Set}

Several existing public traffic datasets include flow features datasets, as well as raw traffic datasets. For instance, KDD CUP1999 and NSL-KDD have forty-one predetermined features in their datasets [19]. Such datasets cannot fulfill our specific requirements for raw traffic, since these data sets contain few VPN and regular traffic samples. However, ISCX VPN-nonVPN dataset [3] comprises of six features of encrypted traffic and six features of other network protocol that it solves our concern about raw traffic.

In comparison, we collected and utilized traffic from an actual traffic dataset. There are six traffic classes in this dataset, including chat, email, FTP, multimedia streaming, VoIP, and VPN. The utilization of such data sets results in finding the best model for classifying traffic and validating it compared to classification aims.
The specific content characteristics of this data set is shown in Table \ref{table:1}.

   \begin{table}[t]
        \centering
        \begin{tabular}{||c|c|c|c||}
            \hline
            Traffic Type & Content & Labeled No. & $\#$ of samples \\ [0.6ex]
            \hline
            1 & Chat & '0' & 22626\\
            \hline
            2 & Email & '1' & 14563\\
            \hline
            3 & Ftp & '2' & 12362\\
            \hline
            4 & Streaming & '3' & 16882\\
            \hline
            5 & VoIP & '4' & 13780\\
            \hline
            6 & VPN & '5' & 54537\\
            \hline
        \end{tabular}
        \caption{Structure of dataset and their label.}
        \label{table:1}
    \end{table}

\subsection{The Proposed Method}
 In our proposed method, we classify non-VPN flows based on their type of application, and VPN is classified as a kind of flow that does not fit to any application. Two metrics, namely, class scores and distance from the class center are used to classify each traffic flow. In the first phase, traffic flows are classified based on their score regarding each class of application, and VPN is the kind of flow that does not get the minimum score required. In the second phase, a one-hot representation is used to represent every individual class center, then each traffic flow is assigned to the corresponding class based on their distance from these centers; the VPN traffic flows' distance is more than maximum permitted distance.

\subsubsection{Score Method}

     The last layer of each neural network is usually devoted to the task of classification. In this way, the number of neurons in this layer is usually equal to the number of classes to be classified, and input is assigned to a class that its corresponding neuron has the maximum score. If there are $m$ classes available, and the corresponding score of each neuron in the last layer is denoted by $y_{i}$, the assigned class $i^*$ is determined based on the following rule:
    \begin{equation}
        i^* = \argmax_{i=1..m} y_{i}.
    \end{equation}
    Parameter $\lambda$ acts as a threshold for the task of classification in the way that input is only assigned to a class that has the score more than the parameter $\lambda$ based on proposed method for rejection in [11]. Otherwise, it is rejected:
    \begin{equation}
        y_{i}^* = \max_{i=1..m} y_{i}, \; \; \; \; \;
        i^* =
        \begin{cases}
        \text{arg} \ y_{i}^*, & \text{if} \ y_{i}^* > \lambda \\
        \text{rejected}, & \text{otherwise}
        \end{cases}
        .
    \end{equation}
    The VPN traffic is a type of traffic which its class in the decision-making rule is rejected and cannot fit to other traffic types.
 
    The last layer of neural network models, contains five neurons corresponding to each non-VPN traffic classes. The decision-making rule becomes as follows:
    \begin{equation}
        y_{i}^* = \max_{i=1..5} y_{i},  \; \; \; \; \;      
        i^* =
        \begin{cases}
        \arg \ y_{i}^*, & \text{if} \ y_{i}^* > \lambda \\
        \text{VPN}, & \text{otherwise}
        \end{cases}
        .
    \end{equation}

    The classification method is designed in the two phases which consist of neural network model to make the model more precise. The first phase distinguishes VPN traffic from non-VPN ones, and the second phase classifies the non-VPN traffics based on their application. VPN traffic and non-VPN traffic are classified based on the proposed method and their scores, respectively.
    Since the first network also classifies the traffic, it is more efficient to use the first network's information for late classification. In order to do so, parameter $\mu$ which acts as a threshold to assign each input's corresponding class is defined. If the maximum score of the first network exceeds $\lambda$ and is more than $\mu$, the class related to it is chosen as a result. Conversely, if its value is lower than $\mu$, the second network decides about the class of the input data.
    The parameter $\mu$ should be higher than the parameter $\lambda$, and the decision process is as follows:
     \begin{equation}
    y_{i}^* = \max_{i=1..5} y_{i}.\; \; \; \; \;
    \gamma_{i}^* = \max_{i=1..5} \gamma_{i},
    \end{equation}
    \begin{equation}
    	i^* =
    	\begin{cases}
    	\text{arg} \ y_{i}^*, & \text{if} \ y_{i}^* > \mu \\
    	\text{arg} \ \gamma_{i}^*, & \text{if} \ \mu >  y_{i}^* > \lambda \\
    	\text{VPN}, & \text{if} \ y_{i}^* < \lambda
    	,
    	\end{cases}
    \end{equation}

    \noindent where $y_{i}$ and $\lambda_i$ are the class scores of classes produced by the first and the second network, respectively.

\subsubsection{Distance Method}
    Comparing distance of the last layer's values from each class centre by utilizing one-hot representation as to their centers is an effective way of classifying traffic data. Due to the number of available classes, $m$, an $m$-dimensional space is required to assign each class's corresponding one-hot representation as a center, $c_i$. 
    In an $m$-dimensional space, each class center resides on the value of one on each dimension, which means the first class center has its first dimension value equal to one and the other dimensions equal to zero. This rule applies to all classes. The decision-making process treats the values of $m$ neurons of the last year of the model, as a point in an $m$-dimensional space, $z$. At first, the distance of the resulted point from all class centers is computed, and then the corresponding class is the one that its related center has the minimum distance from the resulted point. Thus, the decision-making process can be summarized as
    
    \begin{equation}
        i^* = \argmin_{i = 1..m} \dis(\mathbf{y}, \mathbf{c_i})\;.
    \end{equation}
    \noindent where $\Vec{y}$ and $c_{i}$ are neurons output and class centers, respectively.
    Similarly to $\lambda$, parameter $\eta$ is defined as a threshold for the task of classification in this method:
    \begin{equation}
        d_i^* = \min_{i=1..m} \dis(\mathbf{y}, \mathbf{c_i}) ,  \; \; \; \;
        i^* =
        \begin{cases}
        \arg \ d_{i}^*, & \text{if} \ d_{i}^* < \eta \\
        \text{rejected}, & \text{otherwise}
        \end{cases}
        .
    \end{equation}

    After applying each raw input traffic to the model, a point in a five-dimensional spaces is found. The classification task is completed based on the following decision rule:
    \begin{equation}
        d_i^* = \min_{i=1..5} \dis(\mathbf{y}, \mathbf{c_i}) , \; \; \; \;
        i^* =
        \begin{cases}
        \arg \ d_{i}^*, & \text{if} \ d_{i}^* < \eta \\
        \text{VPN}, & \text{otherwise}
        \end{cases}
        .
    \end{equation}
    
    Similar to score method procedure, two-phase classification consisting of two networks that are used for this technique. The first network distinguishes VPN traffic from non-VPN ones, and the second network classifies non-VPN traffics based on their applications. Likewise to score method procedure, the parameters $\delta$ and $\eta$ are used to act as a threshold to choose the type of phase for classification and distance baseline. Furthermore, the parameter $\delta$ should be less than the parameter $\eta$, and the decision-making process is as follows:

     \begin{equation}
        d_i^* = \min_{i=1..5} \dis(\mathbf{y}, \mathbf{c_i}) ,\; \; \; \;
    	d_{2i}^* = \min_{i=1..5} \dis(\mathbf{\gamma}, \mathbf{c_{2i}}) ,
    \end{equation}
    \begin{equation}
    i^* =
    \begin{cases}
    \arg \ d_{i}^*, & \text{if} \ d_{i}^* < \delta \\
    \arg \ d_{2i}^*, & \text{if} \ \delta <  d_{i}^* < \eta \\
    \text{VPN}, & \text{if} \ d_{i}^* > \eta

    \end{cases}
    ,
    \end{equation}
    where $c_{2i}$ are the classes centers in the second network, and $\gamma$ is the output of the second network.

\section{Models}
    In order to evaluate our proposed method of classification, we used the decision-making rule on two different models of neural networks.
\subsection{MLP}
    The suggested cascade neural network model for the first phase, is a three-layer linear network using a particular non-linear activation function for each layer. 
    To perform the task of feature extraction, the weights of neurons ($W_{784\times1000}$) in the first layer are multiplied by the chosen features from each flow, which has 784 dimensions ($x_{ 784\times1}$). In this approach, the model attempts to expand the space to new dimensions to discover correlations between them, so the dimension of weights of neurons should be 784 $\times$ 1000, which means that output of the first layer ($q_{1000\times1}$) dimension is 1000 and is computed as $q = Wx$.

    The suggested activation function for the first and second layers is ReLU, which is defined as 
    \begin{equation}
        ReLU(x) = \max(0,x).
    \end{equation}
    As a result, the outputs of the first layer, $s_{1000\times1}$, the second layer, $r_{1000\times1}$, and $p_{1000\times1}$ which is the second layer neurons to project the first layer results to the lower dimension will be obtained as follows, where $U_{1000\times100}$ is the weight matrix of the second layer:
    \begin{equation}
        s = ReLU(q),
    \end{equation}
    \begin{equation}
        p = Us, \; \; \; \; \;  r = ReLU(p).
    \end{equation}
    
    As for the last layer, the number of neurons should be matched with the number of classes, excluding VPN traffic. Then, the weight matrix in this layer is $V_{100\times5}$, and the output after multiplication is $z_{5\times1}$ as $z = Vr$. The last layer activation function to produce non-linearity is the Gaussian activation function, which is defined as:
    \begin{equation}
        \text{Gaussian}(x) = \exp \ \{{\frac{-||x - c||^2}{2\sigma^2}}\},
    \end{equation}
    \noindent where, for simplicity, we assume that $\sigma^2 = 1$ and $c$ is equal to zero.

    The result of the model will be $y_{5\times1}$, which is computed by applying Gaussian function on each dimension of $z_{5\times1}$:
    \begin{equation}
        y = \text{Gaussian}(z) = \exp \ \{ \frac{-z^2}{2} \},
    \end{equation}
    The model is summarized in the Table \ref{table:2}.
    \begin{table}[t]
        \centering
        \begin{tabular}{||c|c|c|c||}
            \hline
            Layer & Operation \& non-linearity & Input Size & Output Size \\ [0.5ex]
            \hline
            1 & Linear + ReLU & 784*1 & 1000*1 \\
            \hline
            2 & Linear + ReLU & 1000*1 & 100*1 \\
            \hline
            3 & Linear + Gaussian & 100*1 & 5*1 \\
            \hline
        \end{tabular}
        \caption{Structure of network in MLP model.}
        \label{table:2}
    \end{table}

\subsection{LSTM}

LSTM, which is an artificial recurrent neural network, consists of a cell and three regulators, namely, input, output, and forgotten gates. The cell recognizes values over variable amounts of time, and the three gates monitor information flows into and out of the cell. LSTM networks are well equipped to detect, analyze, and make inferences based on time series data. Besides, LSTM has resolved the bursting and disappearing gradient problems in conventional RNN preparation [19]. These benefits of LSTM allows us to implement the second part of cascade neural network to classify rejected traffic flows.

The equation types for an LSTM unit's forward pass with a forget gate are [20]:
    \begin{equation*}
        f_t = \sigma_g  (W_f x_t + U_f h_{t-1} + b_f)  ,
    \end{equation*}
    \begin{equation*}
        i_t = \sigma_g  (W_i x_t + U_i h_{t-1} + b_i) ,
    \end{equation*}
    \begin{equation*}
        o_t = \sigma_g  (W_o x_t + U_o h_{t-1} + b_o) , \; \; h_t = o_t \circ \sigma_h (c_t) ,
    \end{equation*}
    \begin{equation*}
        c_t = f_t \circ c_{t-1} + i_t \circ \sigma_c (W_c x_t + U_c h_{t-1} + b_c) ,
    \end{equation*}
    \begin{equation*}
        h_t = o_t \circ \sigma_h (c_t) ,
    \end{equation*}

\noindent where $c_0 = 0$ and $h_0 = 0$ are the initial values, besides, subscript $t$ and $\circ$ show the time and Hadamard product. $\sigma_g $, $\sigma_c$, and $\sigma_h$  are sigmoid function, hyperbolic tangent function, and hyperbolic tangent function or $\sigma_h (x) = x$, respectively.
Matrices $W$ and $U$ include the weights of the input and recurrent links, respectively [19].

For the second phase of the cascade neural network, a three-layered neural network model was developed. LSTM, the proposed neural network's first layer, receives the first 784 bytes of data flow and treats them as a sequence of values throughout this phase. The output of the LSTM layer, which has a dimension of 300, is then transferred to a 100-dimension output using a linear layer with 100 neurons and the ReLU function (non-linearity). For the last layer, a linear layer with a Gaussian activation function is employed to not only establish non-linearity but also to match the number of neurons to the number of classes. The model structure is summarized in the Table \ref{table:3}.
   \begin{table}[t]
        \centering
        \begin{tabular}{||c|c|c|c||}
            \hline
            Layer & Operation \& non-linearity & Input Size & Output Size \\ [0.5ex]
            \hline
            1 & LSTM & 784*1 & 300*1 \\
            \hline
            2 & Linear + ReLU & 300*1 & 100*1 \\
            \hline
            3 & Linear + Gaussian & 100*1 & 5*1 \\
            \hline
        \end{tabular}
        \caption{Structure of network in LSTM model.}
        \label{table:3}
    \end{table}

\subsection{Proposed End-to-end Learning Framework}
Because of the drawbacks of divide-and-conquer machine learning approach, we propose an E2E learning approach which is shown in Figure 1. Preprocess, training, and validation phases, as well as a test phase, comprise our proposed E2E encrypted traffic classification mechanism. In contrast to classic divide-and-conquer machine learning approaches, this framework does not have distinct modules like feature extraction, feature selection, and classifier. Indeed, the suggested cascade neural network model includes these components. The features are automatically learnt, and the traffic is immediately classified through the last layer of each phase, accomplishing the aim of E2E learning. 

\begin{figure}[h]
\includegraphics[width=8.5cm]{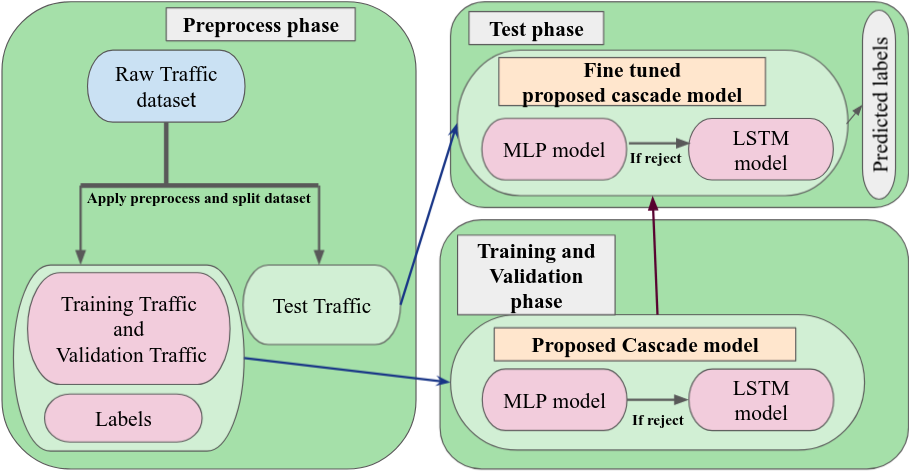}
\centering
\caption{Proposed end-to-end learning framework}
\label{fig:fig}
\end{figure}

The preprocess phase splits the required data for the suggested model's input data from the dataset's raw traffic. The training and validation phase is intended to not only learn the model but also to ensure that it is being learned by comparing loss and metrics' values for training and validation datasets. In addition, Hyper-parameter values, such as learning rate, will also be found in table IV in the Results section. 

The data sample that was utilized in the test phase to provide an unbiased evaluation and generalization of a final model fitted on the training dataset.
The metrics values are the average of metrics for the test dataset that has been trained and validated using 10-fold cross-validation.

\subsection{Loss Functions}
When designing and setting our proposed model, we must specify a loss function, since neural networks are trained via stochastic gradient descent. Two loss functions are employed to configure our suggested model: the Score method loss and the Distance method loss. 

\subsubsection{Score Method Loss}
The score method loss is based on mean square error, which means that the primary objective of this loss is to minimize the squared difference between the model's final result, $y_i$, and the desired output, $s_i$, as follows: 

\begin{equation}
	Loss = ||y_i - s_i ||^2 .
\end{equation}

\subsubsection{Distance Method Loss}

The distance is determined to complete the classification task, as described in the distance method loss.
The model should make the outcome as close to the correct class center as feasible, while keeping it as far away from other classes as possible. Similar to [10], we utilize the loss function of: 

\begin{equation}
	Loss = \sum_{i = 1}^{N} \Bigg (d_{y_i} \Big (x^{(i)} \Big) + \sum_{j \not\in y_i} \max \Big ( 0, \eta - d_j(x^{(i)}) \Big) \Bigg),
\end{equation}
\noindent where $\eta > 0 $, $d_{y_i}  (x^{(i)} )$ is the distance from the correct class, and $d_{j}  (x^{(i)} )$ is the distance from the $j_{th}$ class.
\section{Results}

We separate the first 784 bytes of each network flow as inputs, which comprise multiple traffic packets to train and test our proposed model.
The preprocess, training, validation, and test stages are shown in Figure 1 as described in the E2E learning framework. 

The proposed model is developed using the PyTorch [21] software framework. Table \ref{table:4} shows the values of the model parameters. 

\begin{table}[h]
    \centering
    \begin{tabular}{||c|c|c|c||}
        \hline
        Parameters & Batch Size & Learning Rate & Decaying LR weight\\ [0.4ex]
        \hline
        Value & 64 & 0.0001 & 0.05  \\
        \hline
    \end{tabular}
    \caption{Proposed method learning parameters}
    \label{table:4}
\end{table}

Adam optimizer, as an optimization algorithm, is utilized to update network weights iteratively based on training data, because of the desirable properties of Adam optimizer. For instance,  hyper-parameters have intuitive interpretation and typically require little tuning and it is appropriate for non-stationary objectives, as mentioned in [22].

Furthermore, both validation size and test size are 0.2 of the dataset, allowing for a more thorough evaluation of the proposed model's performance. Hyper-parameter tuning yields the value of threshold hyperparameters, which can be found in Table \ref{table:5}, and are the same for the two mentioned loss functions. 

\begin{table}[h]
    \centering
    \begin{tabular}{||c|c|c|c|c||}
        \hline
        Hyperparameter & $\mu$ & $\lambda$ &  $\eta$ & $\delta$\\ [0.5ex]
        \hline
        Value & 0.85 & 0.70 & 0.40 & 0.30 \\
        \hline
    \end{tabular}
    \caption{Rejection (threshold) hyperparameters}
    \label{table:5}
\end{table}

\subsection{Performance Metrics}
Accuracy, precision, recall, and F1-score are the four evaluation metrics utilized. The general performance of a classification model is assessed by its accuracy. Each class of traffic's performance is examined using precision, recall, and F1-score. The formulas for each metric are described below:

\begin{equation*}
	Accuracy = \frac{TP + TN}{TP + TN + FP +FN} ,
\end{equation*}

\begin{equation*}
	Precision = \frac{TP}{TP + FP},\; \; Recall = \frac{TP}{TP + FN}  
\end{equation*}

\begin{equation*}
	F1-score = \frac{2 * Precision * Recall}{Precision + Recall}.
\end{equation*}

\noindent where TP (True Positive) and TN (True Negative) are results in which the model correctly predicts the positive and negative classes, respectively. In contrast, FP (False Positive) and FN (False Negative) are the results of the model erroneously predicting the positive and negative classes, respectively.  

\subsection{Model Performance}
In this section, the performance metrics of the proposed model for the two described loss functions are presented.
Admittedly, the model has been trained with the hyperparameters listed in table \ref{table:4} and table \ref{table:5}, besides, performance metrics for the test dataset will be reported. 

\begin{table}[h]
    \centering
    \begin{tabular}{||c|c|c|c|c|c|c||}
        \hline
        Traffic Type & Chat & Email &  FTP & Streaming & VoIP & VPN \\ [0.5ex]
        \hline
        Precision & 0.80 & 0.79 & 0.81 & 0.90 & 0.89 & 0.88 \\
        \hline
        Recall & 0.78 & 0.78 & 0.90 & 0.88 & 0.86 & 0.87 \\
        \hline
        F1-Score & 0.79 & 0.78 & 0.85 & 0.89 & 0.87 & 0.87 \\
        \hline
    \end{tabular}
    \caption{Performance metrics for score method}
    \label{table:6}
\end{table}

\begin{table}[h]
    \centering
    \begin{tabular}{||c|c|c|c|c|c|c||}
        \hline 
        Traffic Type & Chat & Email &  FTP & Streaming & VoIP & VPN \\ [0.5ex]
        \hline
        Precision  & 0.83 & 0.81 & 0.85 & 0.94 & 0.93 & 0.94 \\
        \hline
        Recall  & 0.81 & 0.80 & 0.93 & 0.92 & 0.89 & 0.93 \\
        \hline
        F1-Score  & 0.82 & 0.80 & 0.88 & 0.92 & 0.91 & 0.93 \\
        \hline
    \end{tabular}
    \caption{Performance metrics for distance method}
    \label{table:7}
\end{table}

In addition, the accuracy of the distance method and the score method is 0.95 and 0.90, respectively. When the results of performance metrics for two methods are compared, the distance method achieves better results and is used to compare with state-of-the-art models. 

\subsection{Model performance comparison}
The purpose of this section is to compare the results of our proposed method (Cascade-NN) to those of state-of-the-art models based on E2E [6]. Accuracy, precision, recall and F1-score are the metrics for models evaluation. 

\begin{table}[h]
    \centering
    \begin{tabular}{||c|c|c|c|c|c|c||}
        \hline
         & Precision & Recall & F1-Score & Accuracy \\ [0.5ex]
        \hline
        Cascade-NN & 87.8 & 87.6 & 87.3 & 94.0  \\
        \hline
        EE-CNN & 86.8 & 87.2 & 86.9 & 86.0 \\
        \hline
        Improvement & 1.0 & 1.4 & 0.4 & 8.0 \\
        \hline
    \end{tabular}
    \caption{Comparison of performance metrics ($\%$) }
    \label{table:8}
\end{table}

In summary, our proposed method outperforms the current state-of-the-art method based on E2E learning. It validates the efficacy of our current proposed end-to-end encrypted traffic classification method based on a cascading neural network by rejection. Despite of the fact that the number of trainable parameters in CNN depends on the input image size and parameters' value, its complexity is too large when in comparison with our proposed method.

\section{CONCLUSIONS}

A novel end-to-end encrypted traffic classification approach utilizing deep neural networks was presented in this article focused o n the study of a conventional encrypted traffic classification approach utilizing a divide-and-conquer technique. The approach combines feature configuration, extraction of features, and compilation of features into a common structure. Therefore, it can obtain further traffic features efficiently. Contrary to either the divide-and-conquer approach and other strategies of machine learning, the end-to-end approach has a strong adaptive impact. We noticed that the proposed neural networks are far quite suited than prior machine learning solutions to the challenge of encrypted traffic classification. The results on the mentioned datasets brought substantial refinements to all of the state-of-the-art methods, confirming the reliability of our envisaged end-to-end principle. Recent research has shown that deep learning techniques, including cascade neural networks by rejection have excellent prospects in the traffic classification area. We intend to accurately analyze the solution suggested in this article to enhance classification of traffic capabilities.




\end{document}